 \documentclass[amsmath,amssymb,showpacs]{revtex4}
\usepackage{amsfonts}    
\usepackage{amssymb}
\usepackage{latexsym}
 \usepackage{graphicx}
 \usepackage{rotating}
\usepackage[usenames,dvipsnames]{color}
\usepackage[active]{srcltx}

\renewcommand{\qquad}{\hspace*{25pt}}

\def\pt(#1){({\it #1\/})}

\newcommand{\ms}{\noalign{\vspace{3\p@ plus2\p@ minus1\p@}}}
\newcommand{\bs}{\noalign{\vspace{6\p@ plus2\p@ minus2\p@}}}

\begin{document}

\title{Five-body choreography on the algebraic lemniscate is a potential motion}

 \author{Juan Carlos Lopez Vieyra}

\email{vieyra@nucleares.unam.mx}
\affiliation{Instituto de Ciencias Nucleares, Universidad Nacional
Aut\'onoma de M\'exico, Apartado Postal 70-543, 04510 M\'exico,
D.F., Mexico{}}

\begin{abstract}
In a remarkable paper of 2003 by Fujiwara et al. \cite{Fujiwara2003},   
a figure-eight  three-body choreography on the algebraic lemniscate by Bernoulli was discovered. 
Such a choreography was found to be driven by the action of a pairwise potential $ V(r_{ij}) $,  
depending only on the mutual relative distances $r_{ij}, i,j=1,2,3$.
In the present Letter we show that two different choreographies of five bodies on the same 
algebraic lemniscate exist and correspond to solutions of ten coupled Newton equations of motion 
with a pairwise interaction potential. For each choreography the explicit form of the potential is found
and ten constants of motion are presented, thus, it is superintegrable.
 \end{abstract}

\pacs{45.20.Dd, 45.50.Jf}

\maketitle


\section{Introduction}

In 1993, a remarkable figure-eight  trajectory for a choreographic solution of the three-body problem 
under Newton gravity law was discovered numerically by C. Moore \cite{moore}. Later, a rigorous proof 
of its existence was given by A. Chenciner and R. Montgomery \cite{chenAndMont}.  Since then,
many  $N$-body choreographic  trajectories in Newton gravity have been found, all numerically
(see e.g. C. Sim\'{o} \cite{simo1,simo2}). No analytic form of the figure eight trajectory 
(as well as for the other choreographic trajectories) is known so far, although 
it can be approximated with high precision by algebraic curves \cite{simo2}.

Notably, the figure-eight  trajectory in the three-body Newton problem discovered by Moore looks very much similar
to a well know algebraic lemniscate \cite{simo2}. It seems that such a similarity motivated 
Fujiwara et al. \cite{Fujiwara2003} to study a three-body choreography on the algebraic lemniscate by J. Bernoulli. 
They found that such a  choreographic solution exists being a solution of the Newton equations under the potential 
$V\sim 1/4\log r_{ij}^2 - \sqrt{8}/3\, r_{ij}^2$ where \( r_{ij} \) is the relative distance between the bodies $i$ and $j$ 
which lie on a common plane \cite{Fujiwara2003}.
Such potential actually corresponds to a Newtonian gravitational force in two dimensions plus a mutual 
repulsive force of an inverted harmonic oscillator potential.

In order to study the three body choreography on the lemniscate, the trajectory was parametrized by elliptic Jacobi functions.
In particular, in \cite{Fujiwara2003} it was found that such a choreography exists only if the period $\tau $
(or ``equivalently'' the elliptic modulus of the Jacobi functions) takes a specific, concrete value.
Several constants of motion along the trajectory were found indicating that such trajectory is superintegrable.
In a subsequent paper \cite{Fujiwara2004}  the problem of \hbox{$N=5,7$-bodies} on the algebraic lemniscate was investigated.
In each case,  it was found that only for some particular values of the period $\tau$, for which the center of mass remains fixed,
such choreographies are possible: two  different values of the period for the 5-body case and
three different  periods for  the 7-body case were found \cite{Fujiwara2004}. However,  in contrast to the three body case,
no potential for which these choreographies are solutions of the Newton equations of motion was found. It was even conjectured
in \cite{Fujiwara2004} that such choreographies may not satisfy equations of motion under {\em any} interaction potential.

In this Letter, we show that the choreography for $N=5$ bodies on the algebraic lemniscate is a solution
of ten coupled Newton equations of motion with a certain pairwise potential, $V=V(r_{ij})$, depending on the mutual
relative distances between the bodies $r_{ij}$  ($i,j=1\ldots 5$). Such a potential  is found explicitly,
being in a way (see below), similar to the potential for the 3-body choreography, {\it i.e.} a potential of the form
$V\sim \alpha\log r_{ij}^2 - \beta\, r_{ij}^2$ for certain values of the parameters $\alpha, \beta$.
 The question about the maximally particular superintegrability of the closed trajectories
(Turbiner's conjecture~\cite{turbiner2013JPhA}) for both 3-body and 5-body choreographies, is briefly discussed.

\section{Generalities}

The lemniscate of Bernoulli  is a  planar  (8-shaped) algebraic curve of degree 4 satisfying the equation
 \begin{equation}
 \label{eqlemniscate}
 (x^2+y^2)^2 =  c^2 (x^2-y^2)\,,
 \end{equation}
where $c$ is a parameter related to  the homothety of the lemniscate. We parametrize the
algebraic lemniscate by using the Jacobi's elliptic functions $\mathrm{sn}(t,k)$ and $\mathrm{cn}(t,k)$ as
\begin{equation}
\label{xy}
\mathbf{x}(t)=  c\,\frac{\mathrm{sn}(t,k)}{1+\mathrm{cn}^2(t,k)}\,( \hat{\mathbf{x}} + \mathrm{cn}(t,k)\hat{\mathbf{y}}),
\end{equation}
with  $\hat{\mathbf{x}}$ and $\hat{\mathbf{y}}$ being  the Cartesian orthogonal unit vectors on the plane.
In   (\ref{xy})  $k$ stands for  the {\it elliptic modulus} of the Jacobi functions, and the parameter
$t$ is identified with the physical time. Then, parametrization (\ref{xy}) describes  a  motion  on the algebraic 
lemniscate with period $\tau(k)=4\,K(k)$, where $K(k)$ is the complete elliptic integral of the first kind:
\begin{equation}
 \label{ellipticintegral}
 K(k)= \int_0^1 \frac{dx}{\sqrt{(1-x^2)(1-k^2x^2)}}\,.
\end{equation}
The period $\tau(k)$ is defined by the motion on the $x$-direction. In $y$-direction the motion has a period which is
half the period in $x$. The period $\tau$ is $c$ independent. Without loss of generality we can set $c=1$
in the discussion which follows. We restrict the study to  equal mass bodies $m_i=1,\,  i=\ldots {N}$.


\bigskip

 A three body choreography  {\it i.e.}  a periodic motion on a closed orbit, where the bodies chase
each other on a common orbit with equal time-spacing, is defined  on the lemniscate (following the nomenclature in \cite{Fujiwara2003}) by
\begin{equation}
\label{threebodychor}
\{\mathbf{x}_{1}(t), \mathbf{x}_{2}(t), \mathbf{x}_{3}(t)\}
=\{\mathbf{x}(t), \mathbf{x}(t+\tau/3), \mathbf{x}(t-\tau/3)\},
\end{equation}
where $\mathbf{x}(t), \mathbf{x}(t\pm \tau/3)$ are given  in (\ref{xy}).
The center of mass should be fixed (at the origin):
${\mathbf x}_{\scriptstyle \sc CM}=\sum_{i=1}^3 \mathbf{x}_{i}(t)= 0$. This condition
is satisfied if and only if the elliptic modulus $k$ takes the value
\begin{equation}
\label{k0}
k_0^2=\frac{2+\sqrt{3}}{4},
\end{equation}
(see \cite{Fujiwara2003} for details).

\medskip

In the study carried out in  \cite{Fujiwara2003}  a number of constant quantities for the 3-body choreography was found.
In particular, the angular momentum $\sum_{i=1}^3 {\mathbf x}_i \times {\mathbf v}_i = 0$, and
the moment of inertia $\sum_{i=1}^3 {\mathbf x}_i^2=\sqrt{3}$,  are constant.
Some of these conserved quantities are global integrals of motion in involution
(in the sense of a vanishing Poisson bracket), while some others are conserved quantities only along the 
lemniscate trajectory. Conserved quantities along special
trajectories are called generically  ``particular integrals'' \cite{turbiner2013JPhA}.

\medskip

Among the conserved quantities in the 3-body choreography, it was discovered  that
the total kinetic energy  is a constant of motion {$T=1/2\sum_{i=1}^3 {\mathbf v}_i^2 = 3/8$}.
Thus, for a conservative system it immediately implies  that the potential energy is also a constant of motion.
This fact is of a crucial importance to find the form of the potential supporting the choreography of
3 and 5 bodies on the lemniscate.  Since our approach is essentially different from that
used in \cite{Fujiwara2003} to find the potential, we analyze first the case of the three-body  choreography:

\medskip

In the case of the three-body choreography found in \cite{Fujiwara2003} two momentum-independent conserved
quantities were found:
\begin{align}
\label{IntegralsI13b}
I_1 & \equiv  r_{12}^2 \,r_{13}^2 \,r_{23}^2 = \frac{3\sqrt{3}}{2},\\
\label{IntegralsI23b}
I_2 &\equiv  r_{12}^2 + r_{13}^2 + r_{23}^2 = 3\sqrt{3}\,,
\end{align}
where \( r_{ij}^2= {( {\mathbf x}_i -{\mathbf x}_j )^2}\,, i,j=1,2,3\) are the relative distances squared between the bodies $i$ and $j$.
It is evident   $r_{ji}^2=r_{ij}^2$. The {\it integral} $I_2$, being the sum of squares of all relative distances, is denoted as
{\em hyper-radius squared} in the space of relative distances. It appears to play an important role in several problems depending
on relative distances only, {\it i.e.} in translational and rotationally invariant problems.

A natural assumption is that the potential, being a constant of motion, should depend on these momentum-independent conserved quantities,
{\it i.e.}
\begin{equation}
 V= V(I_1,I_2)\,,
\end{equation}
and, if we require a pairwise interaction potential, the explicit  form is
\begin{equation}
\label{V3bgen}
 V= \alpha \log{I_1} + \beta I_2\,.
\end{equation}
where $\alpha$ and $\beta$ are constant parameters which are found by demanding that the Newton equations of motion
are satisfied. In \cite{Fujiwara2003} it was found that  $\alpha=\frac{1}{4},\beta= - \frac{\sqrt{3}}{24}$. Then,
the potential is constant with the value
\(
 V=\frac{1}{4}\log(\frac{3}{2}\sqrt{3}) - \frac{3}{8}\, ,
\)
and the total energy of the three body choreography is
\begin{equation}
\label{E3}
E^{(3)}=T+V= \frac{1}{4}\log(\frac{3}{2}\sqrt{3}) = 0.23869281311055480691\,.
\end{equation}

In total six independent conserved quantities  depending on coordinates and momenta were found
in \cite{Fujiwara2003}. Most of the conserved quantities found in \cite{Fujiwara2003} are {\it particular}
(along the trajectory) integrals of motion.
This fact already indicates that the system  is particularly superintegrable along the lemniscate;
a superintegrable Hamiltonian system is by definition a system with more integrals of motion than degrees of freedom.
It is called maximally superintegrable if there exits $2n-1$ independent integrals of motion for a system with $n$
degrees of freedom. In the case of three particles on the lemniscate we have (after removing two degrees of
freedom corresponding  to center of mass)   $n=4$ degrees of freedom.
If the three body choreography on the lemniscate supports Turbiner's conjecture (2013), namely that
{\em any closed periodic trajectory is  particularly (maximally) superintegrable} there should be one more independent
constant of motion along the trajectory to be discovered. This will be investigated elesewhere.

\section{5-body choreography}
%
\enlargethispage{20pt}
Now consider a five-body choreography on the algebraic lemniscate defined by
\begin{equation}
 \label{fivebodychor}
\{\mathbf{x}_{1}(t), \mathbf{x}_{2}(t), \mathbf{x}_{3}(t), \mathbf{x}_{4}(t), \mathbf{x}_{5}(t) \}
=\{\mathbf{x}(t-2\tau/5), \mathbf{x}(t-\tau/5), \mathbf{x}(t),  \mathbf{x}(t+\tau/5),  \mathbf{x}(t+2\tau/5)\},
\end{equation}
where $\tau=4K(k)$ is the period of the choreography, and the 2-dimensional position vectors
${\mathbf x}(t), {\mathbf x}(t\pm \tau/5),{\mathbf x}(t\pm 2\tau/5) $  are defined as in
(\ref{xy}).

The condition that the center of mass of  the system  is fixed at the origin
\begin{equation}
  {\mathbf X}_{\rm CM}(t) =
  {\mathbf x}_1 + {\mathbf x}_2 + {\mathbf x}_3 + {\mathbf x}_4 + {\mathbf x}_5 =0\,,
\end{equation}
determines, if exists, the period of the motion, or equivalently the value of the elliptic modulus $k$
of the Jacobi functions in the parametrization (\ref{xy}).
We found two solutions for the elliptic modulus $k$
\begin{equation}
 \label{k125bodies}
 k^2= \begin{cases}
0.653 660 413 954 773 213 45 \ldots \equiv k_1^2\\
0.997 643 736 031 613 235 09 \ldots \equiv k_2^2\\
\end{cases}
\end{equation}
c.f.~\cite{Fujiwara2004}. The solutions (\ref{k125bodies}) also ensure (a) the conservation of the angular momentum:
\[
 {\mathbf L}= \sum_{i=1}^5 {\mathbf x}_i \times {\mathbf v}_i = 0 \,,
\]
and (b) the hyper-radius squared (or the  moment of inertia $\sum^5 {\mathbf x}_i^2$):
\begin{equation}
\label{HR5}
 I_{\rm HR}^{(5)}= 5\sum_{i=1}^5 {\mathbf x}_i^2= \sum^5_{i<j} r_{ij}^2 =
 \begin{cases}
 11.995 383 205 775 537 457 \ \mbox{for}\ k_1\\
 17.975 523 091 392 961 251 \ \mbox{for}\ k_2\\
 \end{cases}
\end{equation}

\begin{figure}
\begin{center}
\begin{tabular}{ccc}
& $t=0$ &\\[-40pt]
\includegraphics[width=2.5in]{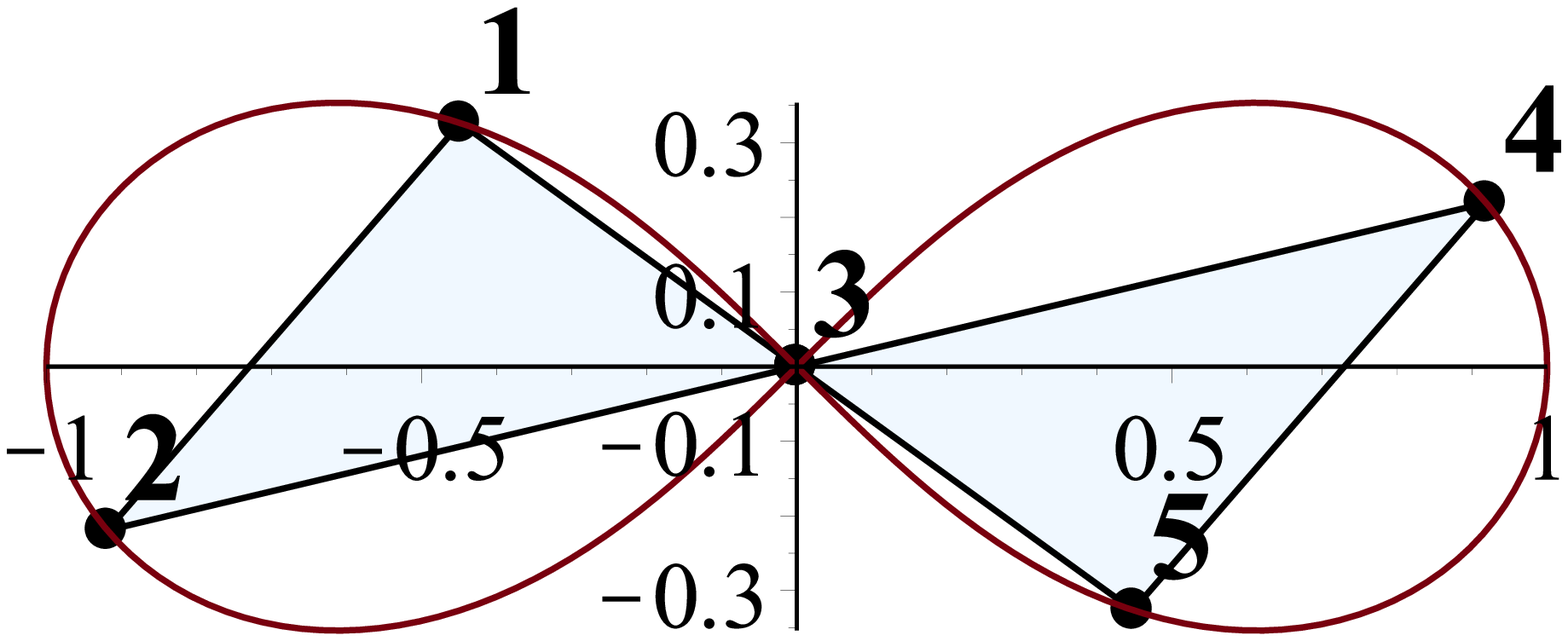} && \includegraphics[width=2.5in]{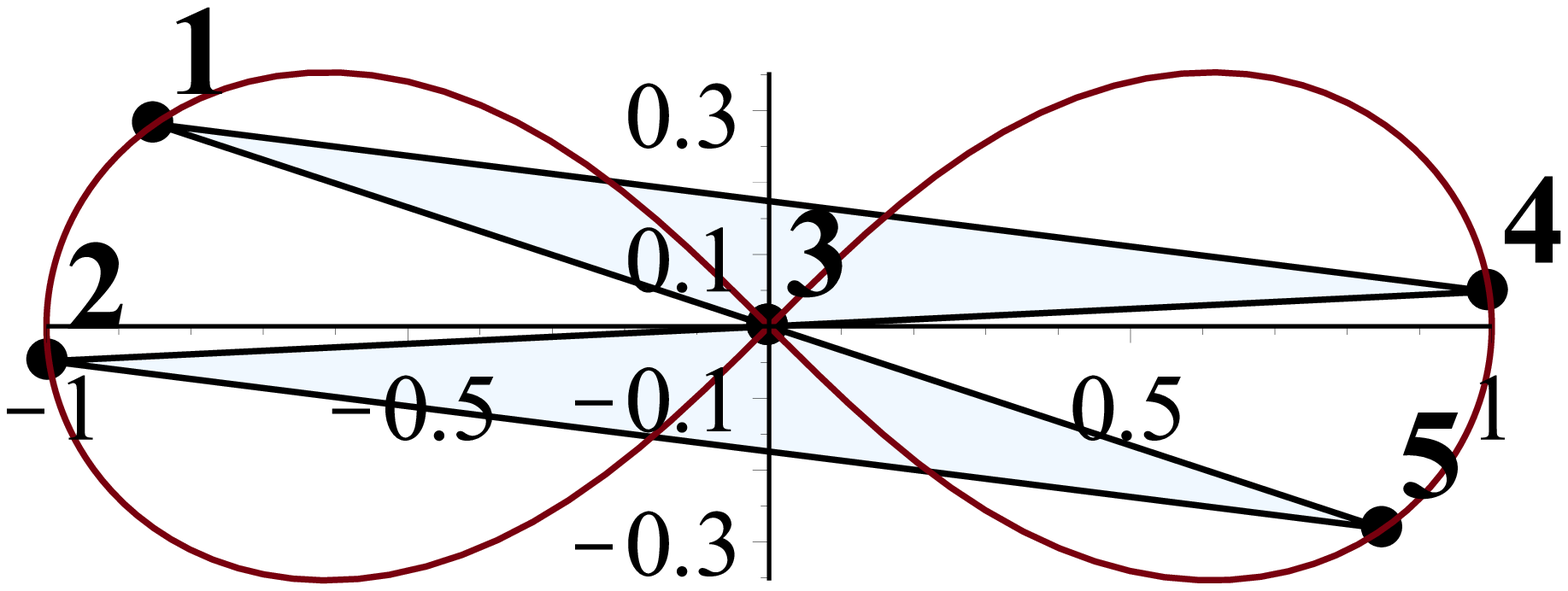} \\[-30pt]
 & $t=K/5$ &\\[-40pt]
 \includegraphics[width=2.5in]{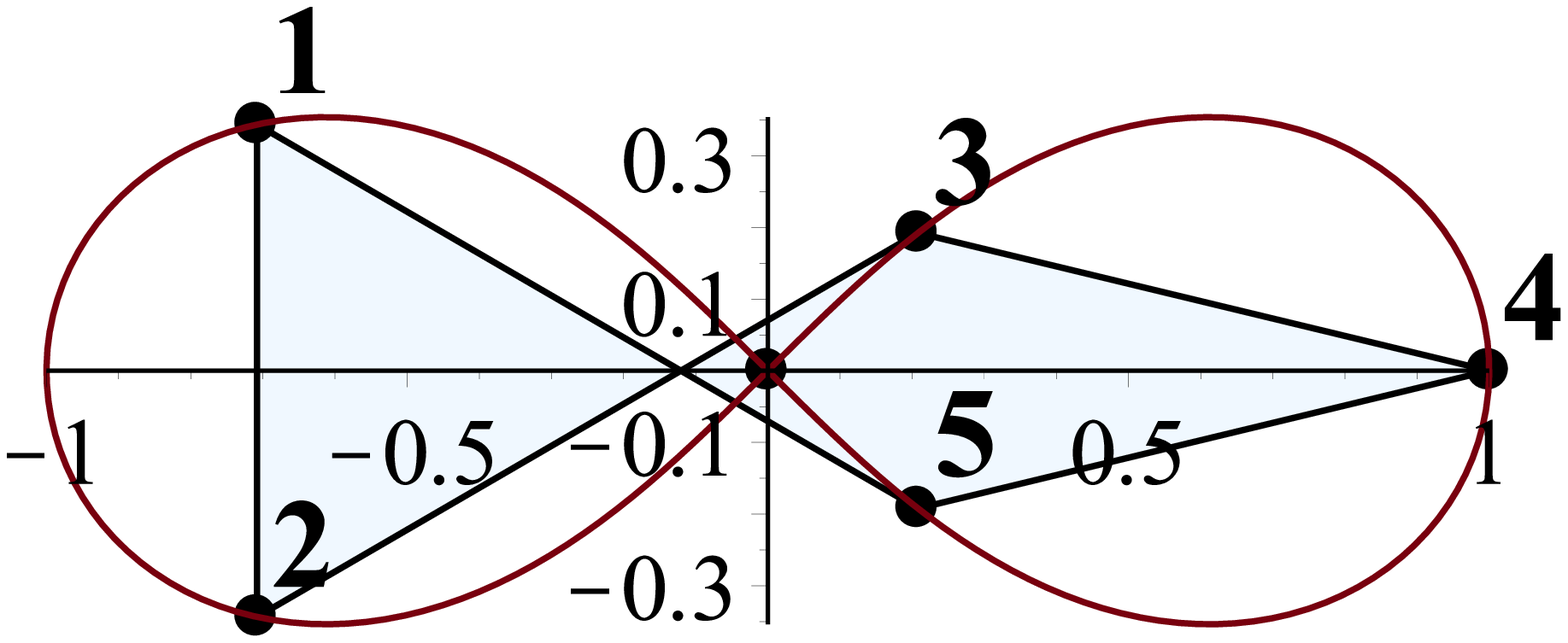} && \includegraphics[width=2.5in]{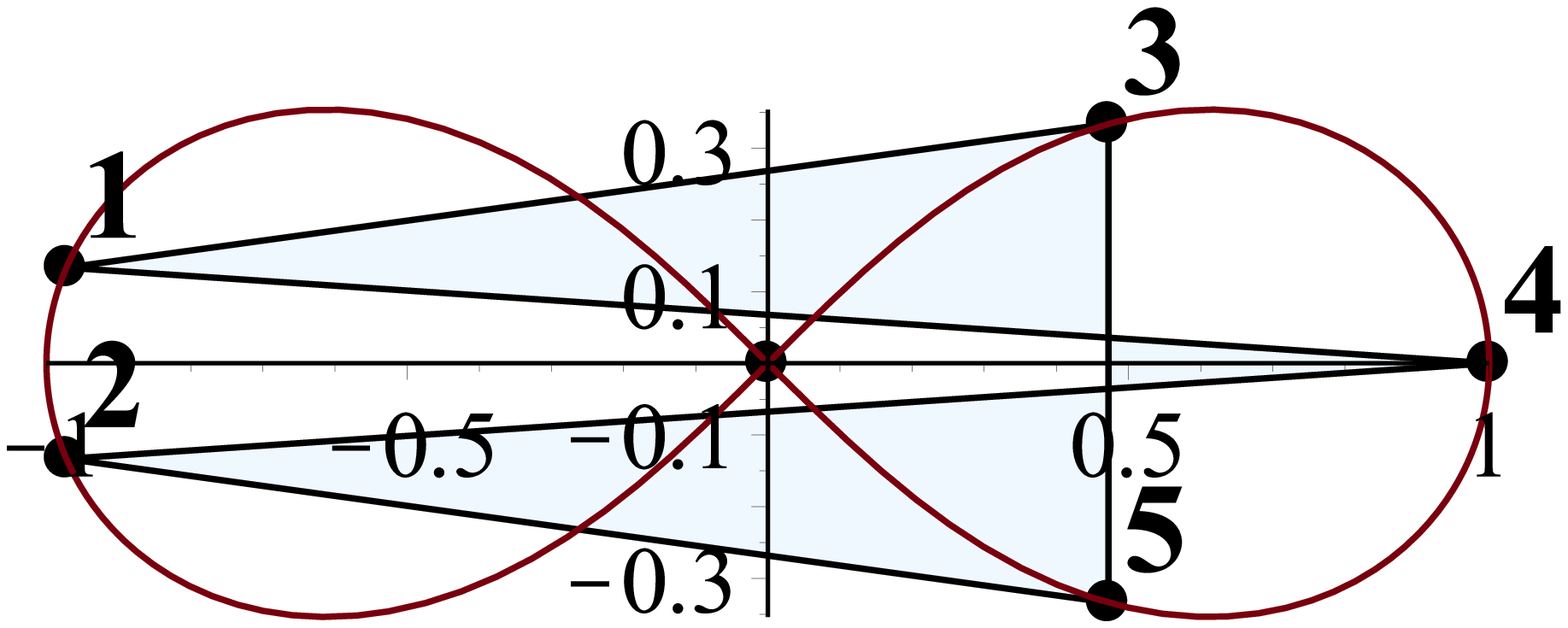} \\[-40pt]
$k^2=k_1^2 = 0.653 660 413 954 773 213 440$ & &$k^2= k_2^2=0.997 643 736 031 613 235 083$ \\[10pt]
 \end{tabular}
\end{center}
\caption{\label{fig1}
The lemniscate and the configuration of five bodies (\ref{fivebodychor}) at $t=0$ and $t=K/5$ for
$k_1^2$ (left) and
for $k_2^2$ (right), see (\ref{k125bodies}). The position of the center of mass is marked by a bullet and placed
at the origin of coordinates. The lines between the bodies indicate the relative distances
involved in the conserved quantities $I_1^{(5)},I_2^{(5)}$ (see the text).
}
\end{figure}

\bigskip

In the search for conserved quantities for the 5-body choreographies on the lemniscate,
we found that in addition to the hyper-radius squared (\ref{HR5}), there exist  two more
momentum independent constants of motion along the algebraic lemniscate:
\begin{equation}
\label{I15}
   I_1^{(5)} =
  \begin{cases}
 r_{12}^2 r_{23}^2 r_{34}^2 r_{45}^2 r_{15}^2  \  \mbox{for} \ k_1\\[7pt]
 r_{13}^2 r_{35}^2 r_{25}^2 r_{24}^2 r_{14}^2   \  \mbox{for} \ k_2\\
\end{cases}
\end{equation}

\begin{equation}
\label{I25}
I_2^{(5)}=
  \begin{cases}
 r_{12}^2 + r_{23}^2 + r_{34}^2 + r_{45}^2 + r_{15}^2   \  \mbox{for} \ k_1\\[7pt]
 r_{13}^2 + r_{35}^2 + r_{25}^2 + r_{24}^2 + r_{14}^2   \  \mbox{for} \ k_2\\
\end{cases}
\end{equation}
To the best of our knowledge such constants of motion for the 5-body choreographies were not known before. 
There are no other constants of motion formed by products or sums of some relative distances squared.

The conserved quantities $I_1^{(5)}, I_2^{(5)}$ depend on
five (out of ten) different relative distances squared $r_{ij}^2$. Moreover, the subsets of relative distances
corresponding to the two possible values of the elliptic modulus $k^2_{1,2}$, are mutually disjoint. Following
the nomenclature (\ref{fivebodychor}), for the
choreography corresponding to the smaller value of the elliptic modulus squared $k_1^2$ the conserved quantities
$I_1^{(5)}, I_2^{(5)}$ depend on relative distances between the nearest-neighbors
$r_{i, i+1},\, \scriptstyle i=1\ldots 5$ ($r_{5,6}\to r_{5,1}$), while for the choreography corresponding
to the larger value of the elliptic modulus $k_2^2$,
these conserved quantities depend on
the relative distances between next-to-nearest neighbors
$r_{i, i+2},\, \scriptstyle i=1\ldots 5$ ($r_{4,6}\to r_{4,1}$, $r_{5,7}\to r_{5,2}$).
Notice that, alternatively to the hyperradius squared (\ref{HR5}) we can as well define (as independent) the quantity
$I_3^{(5)} \equiv I_{\rm HR}^{(5)} - I_2^{(5)}$ which depends on the relative distances
of the next-to-nearest neighbors for $k_1$, and  on the relative distances of the nearest-neighbors for $k_2$.
So, we can build two sets of relative distances: the set of relative distances between nearest-neighbors and the set of
relative distances between next-to-nearest neighbors. Within each set the relative distances oscillate in the same way
except for time delays which are multiples of a fifth of the period. Thus, these two sets can be
characterized by the amplitude of the oscillations. Representative examples are:

\begin{align}
&[r_{12}^{min}(k_1)=0.68667279299905573944, r_{12}^{max}(k_1)
=1.108730495493916550]\,, \label{rk1}
\\
&[r_{13}^{min}(k_1)=0.38411684326397186757, r_{13}^{max}(k_1)
=1.889145301630350845] \nonumber
\end{align}
\begin{align}
&[r_{12}^{min}(k_2)=0.26636437357508228993, r_{12}^{max}(k_2)
=1.7913847571932386574]\,,\label{rk2}
\\
&[r_{13}^{min}(k_2)=0.66719007021760872692, r_{13}^{max}(k_2)
=1.9952819069627494321]. \nonumber
\end{align}
In Fig. \ref{fig1} we show  two representative configurations of the five-body choreography
on the lemniscate at $t=0$ and  $t=K/5$, for the case of the elliptic modulus squared $k^2=k_{1}^2$ (left panel)
and  for $k^2=k_{2}^2$ (right panel). For the case $k_1^2$ (left panel in Fig. \ref{fig1}): at $t=0$ the relative distances $r_{15},  r_{24}$
take their maximal values ($r_{15}^{max} = r_{12}^{max}$, $r_{24}^{max} = r_{13}^{max}$) given in (\ref{rk1}) while at
$t=K/5$ the relative distances  $r_{12}$ and $r_{35}$ take their minimal values ($r_{12}^{min}$, $r_{35}= r_{13}^{min}$)
also given in (\ref{rk1}). On the other side, for the case $k_2^2$ (right panel in Fig.\ref{fig1} ):
at $t=0$ the relative distances $r_{15},  r_{24}$
take their maximal values ($r_{15}^{max} = r_{12}^{max}$, $r_{24}^{max} = r_{13}^{max}$) given in (\ref{rk2}) while at
$t=K/5$ the relative distances  $r_{12}$ and $r_{35}$ take their minimal values ($r_{12}^{min}$, $r_{35}= r_{13}^{min}$)
also given in (\ref{rk2}).

\bigskip

The key point in the search for the potential is the fact that the total kinetic energy  along the lemniscate
 \hbox{\(
T= \frac{1}{2}\sum_{i=1}^5 {\mathbf v}_i^2  \,,
\)}
is a constant of motion for the two choreographies found  (similarly to the case of the three body choreography).
It implies, that the potential energy $V$ is a constant of motion as well.  It naturally suggests  that the potential can be
a function of the momentum-independent constants of motion:
\[V=V(I_1^{(5)},I_2^{(5)},I_{HR})\,.
\]
If we impose the extra condition of a pairwise nature of the interaction potential, it  leads to the explicit form
\begin{equation}
\label{genV5}
 V= \alpha\ \ln I_1^{(5)} +  a\ I_2^{(5)}  -\beta\ I_{\rm HR}^{(5)}\,,
\end{equation}
where the constants $\alpha,a,\beta$ are fixed by requiring that the five vectorial Newton equations of motion where the potential
is unknown
\begin{equation}
\label{newton5b}
 \frac{d^2}{d t^2} {\mathbf x}_i(t)   = -{\nabla_{\mathbf x_i} V}\,\quad i=1\ldots 5
\end{equation}
are satisfied. It leads to a coupled system of  ten first order PDE for the potential $V$.
The solution exists and corresponds to
\begin{equation}
\label{potentials5b}
 V=  \begin{cases}
 { \alpha_1}{ \left\{ \ln r_{12}^2 + \ln r_{23}^2 +\ln r_{34}^2 +\ln r_{45}^2 +\ln r_{15}^2 \right\} }  \\[5pt]
 { \alpha_2}{ \left\{ \ln r_{13}^2 + \ln r_{35}^2 +\ln r_{25}^2 +\ln r_{24}^2 +\ln r_{14}^2 \right\} }
 \end{cases}
 -  { \beta_{1,2}} \sum_{i<j} r_{ij}^2,
\end{equation}
with constants (see (\ref{genV5}))
 \begin{equation}
  \begin{cases}
a=0, \alpha_1 = \frac{1}{4}, \beta_1 = 0.015366041395477321360    \  \mbox{for} \ k_1\\[5pt]
a=0,  \alpha_2 = \frac{1}{4}, \beta_2 = 0.049764373603161323382  \  \mbox{for} \ k_2\,.\\
\end{cases}
\end{equation}
We immediately notice that the parameters $\alpha_{1,2}$ take the same value as the parameter $\alpha$ in thre three-body case (\ref{V3bgen}).
The above potentials correspond to two different choreographies of five bodies on the {\it same} algebraic lemniscate, with periods:
\hbox{$\tau_1=8.0487770522074684844$}  for $k_1$, and \hbox{$\tau_2=17.654582260596687373$} for $k_2$. In the form (\ref{potentials5b}) the potential
appears to depend on all ten relative distances $r_{ij}$. However, it is known that on the plane, the  motion of five bodies is described, in general,
by seven relative distances (in the case of zero angular momentum). The question about how many independent relative distances are needed to
describe the 5-body choreographies on the algebraic lemniscate remains to be clarified.  This will be investigated elsewhere.

\bigskip

Another question of interest is the notion of superintegrability of the orbit. In the following we collect
the constants of motion that we have found for the two choreographies.
First of all, in the case of five bodies on the lemniscate we have (after removing two degrees of freedom corresponding to center
of mass) $n = 8$ degrees of freedom. Let us give the list of conserved quantities (global and particular) found so far,
\begin{enumerate}
 \item $L=\sum_{i=1}^5 {\mathbf x}_i \times {\mathbf v}_i = 0$, Total Angular Momentum
 \item \(E=T+V =
 \begin{cases}
  0.54804692944384581936 \ \mbox{for}\ k_1 \\
  0.31747900688996754830 \ \mbox{for}\ k_2 \\
 \end{cases}
 \) Total Energy
 \item \( I_1^{(5)} =
  \begin{cases}
 r_{12}^2 r_{23}^2 r_{34}^2 r_{45}^2 r_{15}^2 = 0.26362178303408707110  \ \mbox{for}\ k_1 \\
 r_{13}^2 r_{35}^2 r_{25}^2 r_{24}^2 r_{14}^2  =30.760801541637359790  \ \mbox{for}\ k_2 \\
\end{cases}
\)
\item  \(I_2^{(5)}=
  \begin{cases}
 r_{12}^2 + r_{23}^2 + r_{34}^2 + r_{45}^2 + r_{15}^2 = 4.0517817845468308414   \ \mbox{for}\ k_1  \\
 r_{13}^2 + r_{35}^2 + r_{25}^2 + r_{24}^2 + r_{14}^2 = 12.515257719766335417  \ \mbox{for}\ k_2  \\
\end{cases}
\)
 \item \(I_{\rm HR}^{(5)}= 5\sum_{i=1}^5 {\mathbf x}_i^2= \sum^5_{i<j} r_{ij}^2   =
 \begin{cases}
 11.995 383 205 775 537 457 \ \mbox{for}\ k_1\\
 17.975 523 091 392 961 251 \ \mbox{for}\ k_2\\
 \end{cases}
\) Hyper-radius Squared

\item \(
 \sum_{i=1}^5 \rho_i^{-2} =   \frac{9}{5} {I_{\rm HR}^{(5)}} \quad
\) Sum of Squares of Curvature

\item \(
T= \frac{1}{2}\sum_{i=1}^5 {\mathbf v}_i^2 = \begin{cases}
 1.0656784451054396 \  \mbox{for} \ k_1\\
 0.35545935316766729 \  \mbox{for} \ k_2\\
\end{cases}
\) Kinetic Energy

\item $J_i(k) = {\mathbf v}_i^2 + \left(k^2 - \frac{1}{2}\right) {\mathbf x}_i^2 = \frac{1}{2}, \quad  i=1,2,\ldots 5$
 \end{enumerate}
In the list of conserved quantities above, the {\it curvature} is defined as
$\rho^{-1}(t) = \frac{|{\mathbf v}(t) \times{\mathbf a}(t)|}{|{\mathbf v}(t)|^3}$, with
${\mathbf v} = \dot{\mathbf x} \equiv \frac{d}{dt}{\mathbf x}$ is the velocity,
and ${\mathbf a} = \dot{\mathbf v}\equiv \frac{d}{dt}{\mathbf v}$ the acceleration.
For any point on the lemniscate,  the square of the curvature is related to the distance to the origin
\(
\rho^{-2}(t) = 9\,{\mathbf x}^2(t),
\) see \cite{Fujiwara2003}.
Notice that the constants $J_i(k)\, i=1,2,\ldots 5$ which appear to correspond to the motion, in phase space, of five independent
two-dimensional isotropic harmonic oscillators might be only a property of the parametrization of the lemniscate and do not
represent true dynamical relations for arbitray values of the elliptic modulus $k$.
Not all above listed conserved quantities are independent. In total we have ten independent conserved quantities and therefore the
trajectory is particularly superintegrable. If Turbiner's conjecture is valid, {\it i.e.} that a closed periodic trajectory
is maximally particularly superintegrable,  then we should have $2n-1=15$ independent constants of motion for the choreographies of  five bodies
moving  along the lemniscate. In such a case, there are  more independent constants of motion along the lemniscate to be discovered.
This will be investigated elsewhere \cite{TurbinerTBP}.

\bigskip

\section{Conclusion}

We studied a choreographic motion of five bodies of unit mass on the (algebraic) lemniscate of Bernoulli
parametrized by Jacobi elliptic functions.  We found two choreographies with different
periods (two  specific concrete values for the elliptic module)   required for the conservation
of the center of mass of the system  in agreement with the findings of Ref.~\cite{Fujiwara2004}. 
We have found explicitly in total  ten independent (global and particular) constants of motion associated 
to these five body choreographies. Thus, the 5-body choreographies on the algebraic lemniscate are superintegrable.
In particular, it was shown that the total kinetic energyis a conserved quantity. 
Two momentum-independent constants of motion depending on five (out of ten) different
relative distances $I_1^{(5)}, I_2^{(5)} $ along the trajectory were discovered. Also, the hyper-radius squared
$I_{\rm HR}^{(5)}$ in the space of relative distances, which depends on {\it all} ten different relative distances,
is a constant of motion. We have shown that  the potential supporting the choreography of five bodies on the algebraic
lemniscate exists and depends on the momentum-independent constants of motion   $I_1^{(5)}$ and  $I_{\rm HR}^{(5)}$ only. 
 We emphasize that such potential is a solution of ten first order PDE for the potential
$V$ given by the Newton equations. For each choreography, the potential is found explicitly being a function 
of the relative distances with pairwise interactions of the form $V\sim \alpha\log r_{ij}^2 - \beta\, r_{ij}^2$ for certain values of the parameters $\alpha, \beta$.

\bigskip

{\bf Acknowledgments}\\
The author wants to thank A. Turbiner for bringing his attention to the problem and for the numerous valuable discussions, and to C. Simo
for his valuable comments on previous versions of this manuscript.


\bigskip

\begin{thebibliography}{9}



\bibitem{Fujiwara2003}
Fujiwara T, Fukuda H and Ozaki H (2003)  J. Phys. A: Math. Gen. 36 2791



\bibitem{moore}
Moore C (1993) {\it Phys. Rev. Lett} {\bf 70} 3675--3679

\bibitem{chenAndMont}
Chenciner A and Montgomery R (2000)
{\it Annals of Mathematics} {\bf 152} 881--901

\bibitem{simo1}
Sim\'o C  (2001)
{\em The Restless Universe: Applications
of $N$-Body Gravitational Dynamics to Planetary, Stellar and Galactic Systems},
editors B. Steves, J. Maciejewski, NATO Advanced Study Institute, IOP Publishing, Bristol

\bibitem{simo2}
Sim\'{o} C (2002)
{\it Celestial mechanics: Dedicated to Donald Saari for his 60th Birthday.
Contemporary Mathematics} {\bf 292}
(Providence, R.I.: American Mathematical Society)



\bibitem{Fukuda2017}
          Fukuda H,  Fujiwara T and Ozaki H,  (2017)
          J.  Phys A Math:  Gen.  50, 10 105202

\bibitem{Fujiwara2004}
Fujiwara T, Fukuda H and Ozaki H (2004),
(Developments and Applications of Dynamical Systems Theory), 1369, 163-177


%
%
%
%


\bibitem{turbiner2013JPhA}
   Turbiner, A V, (2013)
J.  Phys A Math:  Gen.  025203




\bibitem{TurbinerTBP}
Turbiner A and  L\'opez Vieyra  J C,
work in progress


 \end{thebibliography}
\end{document}